\documentclass[]{article}

\usepackage{graphicx}
\usepackage{subfig}
\usepackage{epstopdf}
\usepackage{epsfig}
\usepackage{authblk}

\usepackage[T1]{fontenc}
\usepackage[utf8]{inputenc}

\title{STEREO observations of HD 90386 (RX Sex):\\a Delta Scuti or a hybrid star?}
\author[1,2]{D.Ozuyar\thanks{dozuyar@ankara.edu.tr}}
\author[2]{I. R. Stevens\thanks{irs@star.sr.bham.ac.uk}}
\author[3]{G. Whittaker\thanks{whittaker@astro.utoronto.ca}}
\author[4]{V. Sangaralingam\thanks{vinothini.sangaralingam@asc-csa.gc.ca}}

\affil[1]{Dept. of Astronomy and Space Sciences, Ankara University}
\affil[2]{School of Physics and Astronomy, The University of Birmingham}
\affil[3]{Physics Department, The University of Toronto}
\affil[4]{School of Physics, The University of Montreal}

\begin{document}

\maketitle

\begin{abstract}

HD 90386 is a rarely studied bright A2V type Delta Scuti star (V = 6.66 mag). It displays short-term light curve variations which are originated due to either a beating phenomenon or a non-periodic variation. In this paper, we presented high-precision photometric data of HD 90386 taken by the STEREO satellite between 2007 and 2011 to shed light on its internal structure and evolution stage. From the frequency analysis of the four-year data, we detected that HD 90386 had at least six different frequencies between 1 and 15 c~d$^{-1}$. The most dominant frequencies were found at around 10.25258 c~d$^{-1}$ (A $\sim$ 1.92 mmag) and 12.40076 c~d$^{-1}$ (A $\sim$ 0.61 mmag). Based on the ratio between these frequencies, the star was considered as an overtone pulsator. The variation in pulsation period over 35 years was calculated to be dP/Pdt = 5.39(2) x 10$^{-3}$ yr$^{-1}$. Other variabilities at around 1.0 c~d$^{-1}$ in the amplitude spectrum of HD90386 were also discussed. In order to explain these variabilities, possible rotational effects and Gamma Dor type variations were focused. Consequently, depending on the rotation velocity of HD 90386, we speculated that these changes might be related to Gamma Dor type high-order g-modes shifted to the higher frequencies and that HD 90386 might be a hybrid star.

\end{abstract}

\section{Introduction}

$\delta$ Scuti stars are located at the intersection of the classical Cepheid instability strip and the main sequence (MS). Their luminosity classes vary between V to III, since some massive samples evolve towards the giant region and across the strip at higher luminosities. Spectral types of these stars range from A2 to F0 on the MS and from A3 to F5 in the giant region \cite{Kurtz}. Also, the mass values are between 1.5 and 2.5 M$_{\odot}$ for $\delta$ Scuties with solar metal abundance and 1.0 -- 2.0 M$_{\odot}$ for metal-poor samples. Therefore, it is known that these stars are either at the core or shell hydrogen-burning stages. They are mostly non-radial pulsators, but some $\delta$ Scuties show radial oscillations. The periods of the pulsations are between 0.02 days and 0.25 days, with a typical amplitude of around 0.02 mag \cite{Breger79}.

HD\,90386 is a rarely studied A2V type, bright $\delta$ Scuti star ($B$ = 6.78 mag, $V$ = 6.66 mag). Its variability was accidentally discovered by Jerzykiewicz \cite{Jerzykiewicz} while it was being used as a comparison star for 23\,Sex. He detected short-term variations with the period of 0.0799 days (A $\sim$ 0.01 mag) in the light curve (LC), and reported that their origins were due to either a beating phenomenon or a non-periodic variation in LC.

\section{{\sl STEREO} Satellite and Data Analysis}

In this paper, we present high-precision photometric data of HD\,90386 taken by the {\sl STEREO} satellite to shed light on its internal structure and evolution stage. During the seasonal observations, 684, 678, 651, and 660 data points were derived from the satellite in 2007, 2008, 2010, and 2011, respectively. This yielded a total of 2,673 points, corresponding to 1,809 hours of observation over four years. The data were provided only from the {\sl HI-1A} instrument on-board the {\sl STEREO-A} satellite. It has been capable of observing background stars with the magnitude of $12^m$ or brighter for a maximum of 20 days and a useful stellar photometer which has covered region around the ecliptic (20\% of the sky) with the field of view (FOV) of $20^{\circ} \times 20^{\circ}$. A more detailed description of the data preparation and background information on {\sl STEREO} can be found in Sangaralingam and Stevens \cite{Sangaralingam} and Whittaker et al. \cite{Whittaker}. The nominal exposure time of the camera is 40 seconds, and putting 30 exposures together on board, a 40-minute integrated cadence has been obtained to transmit for each {\sl HI-1} image (For details of the {\sl HI} instruments refer to Eyles et al. \cite{Eyles}). Therefore, the Nyquist frequency of the data is around 18 c~d$^{-1}$. 

Since the {\sl STEREO} satellite has observed the region quite close to the solar disk, data of HD\,90386 were mostly affected by solar activities. Therefore, the distorted effects were removed from the LCs by using a $3^{rd}$ order polynomial fit, and observation points greater than $3\sigma$ were clipped with the help of a freshly coded pipeline in {\sl IDL}. Four years of seasonal and combined time series were then analysed with the Lomb-Scargle (LS) algorithm. The results found were compared to those of {\sl Period04}. For each periodogram, regional noise levels were determined by averaging the noise values in every 0.5 c~d$^{-1}$, and a specific noise characteristic was thus established. Based on this characteristic, a significance level was calculated with 99\% probability. Frequencies whose amplitudes were greater than this level were then detected. The detection precision in four-year combined data given in Figure~1 was around $10^{-5}$ c~d$^{-1}$ in frequency and $10^{-4}$ mag in amplitude.  \hfill \break

\begin{figure}[!h]
\begin{center}
\includegraphics[width=0.7\textwidth]{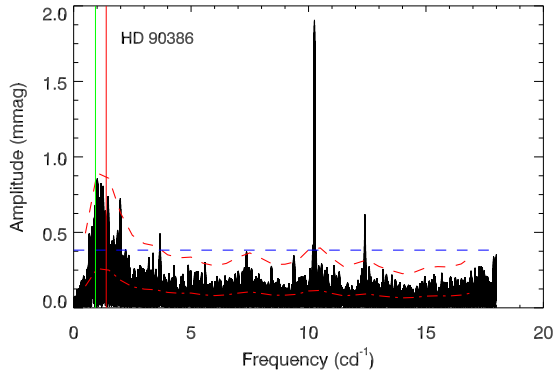}
\caption[Periodogram]{The amplitude spectrum of HD90386 obtained from four-year combined light curves. The mean noise level is plotted with red dash-dot line. Red dashed line is the significance level calculated based on the noise characteristics, and blue dashed line is the 4$\sigma$ level. The estimated rotational periods are shown with green and red vertical lines. For more details, see the text.}
\end{center}
\label{fig:figure1}
\end{figure}

\hfill \break
The noise and related significance levels are presented with red dash-dot and red dashed lines in the figure, respectively. For a comparison, the constant significance level calculated from the mean noise value of the periodogram is shown with a blue dashed line. Additionally, the extracted frequencies, their amplitudes, signal-to-noise ratios (SNR) and noise ($A_m$) values are given in Table~1.

\begin{table}[!t]
\small 
\caption[Frequency analyses results]{The frequencies of HD\,90386, their amplitudes, SNR values, noise levels ($A_m$), and pulsation constants ($Q$).\\}
\begin{center}  
\begin{tabular}{c|c|c|c|c|c}\hline \hline
\multicolumn{6}{c}{\textbf{HD\,90386  Frequencies}} \\\hline
\textbf{No}&	\textbf{Freq.}	&\textbf{Amp.}&\textbf{SNR}&\textbf{$A_m$}&\textbf{$Q$}  \\
&\textbf{(c~d$^{-1}$)}&\textbf{(mmag)}&&\textbf{(mmag)}&\textbf{(days)}\\ \hline
$f_1$	&	10.25258(2)	&	1.92(10)	&	16.88	&	0.11	&	0.031(6)	\\
$f_2$	&	1.99122(5)	&	0.74(10)	&	4.00	&	0.19	&		\\
$f_3$	&	12.40076(6)	&	0.61(10)	&	6.76	&	0.09	&	0.026(5)	\\
$f_4$	&	3.67474(8)	&	0.50(10)	&	4.24	&	0.12	&		\\
$f_5$	&	7.36338(11)	&	0.36(10)	&	3.43	&	0.10	&	0.043(8)	\\
$f_6$	&	9.36921(11)	&	0.34(10)	&	3.92	&	0.09	&	0.034(6)	\\
\hline\hline
\multicolumn{6}{c}{\textbf{2007 Frequencies}} \\\hline
$f_1$	&	10.255(2)	&	1.94(16)	&	9.18	&	0.21	&		\\
$f_2$	&	12.397(6)	&	0.77(16)	&	3.58	&	0.21	&		\\
$f_3$	&	9.371(8)	&	0.59(16)	&	3.74	&	0.16	&		\\

\hline\hline

\multicolumn{6}{c}{\textbf{2008 Frequencies}} \\\hline
$f_1$	&	10.254(3)	&	1.98(17)	&	8.17	&	0.24	&		\\
$f_2$	&	12.413(8)	&	0.65(17)	&	3.59	&	0.18	&		\\

\hline\hline

\multicolumn{6}{c}{\textbf{2010 Frequencies}} \\\hline
$f_1$	&	10.250(3)	&	1.94(20)	&	8.34	&	0.23	&		\\
$f_2$	&	12.399(1)	&	0.55(20)	&	2.93	&	0.19	&		\\

\hline\hline

\multicolumn{6}{c}{\textbf{2011 Frequencies}} \\\hline
$f_1$	&	10.253(4)	&	1.77(22)	&	7.15	&	0.25	&		\\
$f_2$	&	3.666(10)	&	0.69(22)	&	3.21	&	0.21	&		\\
$f_3$	&	12.402(12)	&	0.55(22)	&	3.06	&	0.18	&		\\

 \hline \hline
\end{tabular}
\label{tab:table1}
\end{center}
\vspace*{-0.4cm}
\end{table}

\section{Results and Discussion}
The frequency analysis of the combined data set showed that there were six frequencies having amplitude greater than the specific significance level in the amplitude spectrum (Table~1). As seen in Figure~1, four of these peaks were located between the frequencies of 5 and 15 c~d$^{-1}$. Among them, the most dominant frequency was found at around 10.25 c~d$^{-1}$ (0.0975 days) with the amplitude of 1.92 mmag. Also, the second powerful frequency showed up at 12.40 c~d$^{-1}$ (0.0806 days) in our amplitude spectrum, and it was quite close to the result (0.0799 days) obtained by Jerzykiewicz \cite{Jerzykiewicz}. Based on this alteration in the main frequency, it might be suggested that the star gradually switched pulsation modes from 12.50 to 10.25 c~d$^{-1}$ over time. Furthermore, the intensity of the 12.40 c~d$^{-1}$ in the seasonal LCs seemed to decrease between 2007 and 2011. As a result of this decrement, its amplitude observed in 2010 and 2011 fell below the significance level (Figure~2). However, their SNR values determined from {\sl Period04} were over 4.0, and they were hence included in the analyses. 

\begin{figure}[!t]
\begin{center}
\includegraphics[width=0.78\textwidth]{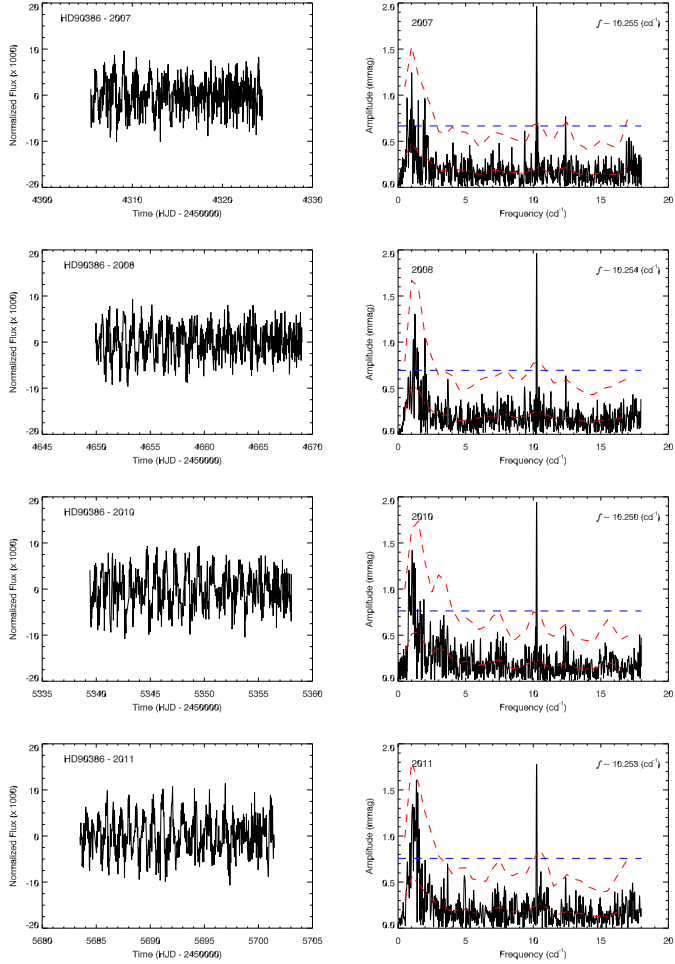}
\caption[Seasonal curves]{Seasonal light curves of HD\,90386 and their amplitude spectra.}
\end{center}
\label{fig:figure2}
\end{figure}

As seen in Figure~2, all seasonal LCs exhibit variabilities around the region of 1.00 c~d$^{-1}$. Since there was not a sufficient number of studies on the physical, chemical, or orbital parameters of the star, interpretation of these variations might not be possible. However, Niemczura et al. \cite{Niemczura} demonstrated that the projected rotational velocities of typical A and F stars ranged from 8 to about 280 km~s$^{-1}$, with a mean of 134 km~s$^{-1}$ in the analysis of high-resolution spectroscopic data of A/F type stars from the Kepler field. Moreover, they found that the maximum of the velocity distribution for A and F stars having an effective temperature between 8,000 and 9,000 K was around 200 km~s$^{-1}$. Assuming that the temperature and radius of the star were around $logT = 3.914$ ($\sim8200$ K; approximated from the $B-V$ colour index) and $R =2.85$ R$_{\odot}$ (calculated from the luminosity equation $L$/L$_{\odot}$ = [$R$/R$_{\odot}$]$^2$ [$T$/T$_{\odot}$]$^4$), its rotational period was estimated to be 0.93(7) and 1.39(11) c~d$^{-1}$ for an inclination angle of $i=90^{\circ}$, and rotational velocities of 134 and 200 km~s$^{-1}$, respectively. These values shown with green and red vertical lines in Figure~1 were considerably compatible with the frequency accumulation at around 1.00 c~d$^{-1}$.

\begin{figure}[!h]
\begin{center}
\includegraphics[width=0.85\textwidth]{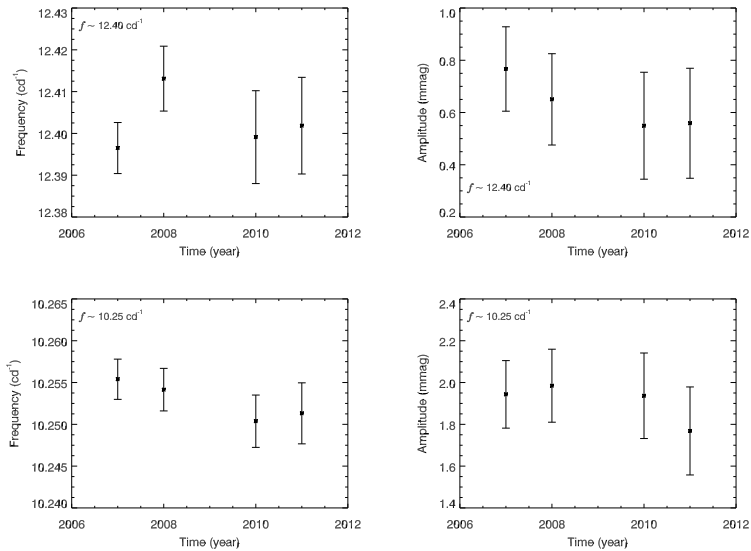}
\caption[Frequency analyses]{Two frequencies of HD\,90386 and their amplitude variations between 2007 and 2011.}
\end{center}
\label{fig:figure3}
\end{figure}

We also computed the pulsation constants ($Q$) of the $\delta$ Scuti type frequencies by using the relation provided by Breger and Bregman \cite{Breger75}. Consequently, $Q$ values of $f_1$ and $f_6$ were around 0.031(6) and 0.034(6) days, which were within the theoretical range for the fundamental mode given by Fitch \cite{Fitch}. If the main frequency was the fundamental mode, its first overtone would be around 13.47 c~d$^{-1}$, which was not observed in our amplitude spectrum. In contrast,  if 9.37 c~d$^{-1}$ was assumed to be the fundamental mode, the first overtone was found as 12.31 c~d$^{-1}$ based on the period ratio $P_1/P_0 = 0.761$. This value slightly matched with $f_3$, and its $Q$ constant was also within the theoretical range given for the first overtone.

Additionally, seasonal frequency and amplitude variations were investigated to gain better understanding of the short-term pulsation behaviour of RX\,Sex. To do this, we determined consistently observed frequencies in seasonal LCs (Figure~2) and surmised that only two frequencies potentially varied as seen in Figure~3. However, all variations were within the error limits, and no significant frequency or amplitude changes were found for both peaks over four years. On the other hand, combining the frequency value (12.5156 c~d$^{-1}$) given by Jerzykiewicz \cite{Jerzykiewicz} with our {\sl STEREO} result, a long-term period variation of $2.49(22) \times 10^{-4}$ yr$^{-1}$ was calculated for the frequency at around 12.40 c~d$^{-1}$. It should also be noted that if there was a gradual change between the frequencies (from 12.50 to 10.25 c~d$^{-1}$), the variation rate of the period would be obtained as $dP/Pdt = 5.39(2) \times 10^{-3}$ yr$^{-1}$ over 35 years. According to Breger and Pamyatnykh \cite{Breger98}, this suggested that the period change observed was not caused by stellar evolutionary changes, but produced by another factor such as non-linear mode interactions.

In order to have a better understanding of the evolutionary stage of HD\,90386, we attempted to estimate some of the fundamental physical parameters. Its effective temperature and luminosity were approximated to be around $logT = 3.914(4)$ from the $B-V$ colour index and $logL/$L$_{\odot} = 1.517(67)$ from the parallax given in {\sl Simbad} Database, respectively. Using these parameters, the radius of the star was derived as 2.85(23) R$_{\odot}$. The surface gravity ($logg=3.887(159)$) was obtained from the relation given by Tsvetkov \cite{Tsvetkov} and the mass ($M/$M$_{\odot}= 2.28(9)$) was computed by means of the equation taken from Cox and Pilachowski \cite{Cox}.

The luminosity and temperature values indicate that HD\,90386 is quite close to the blue edge of the instability strip and to the end of the MS stage where central hydrogen is consumed (Figure~4). Its metallicity index ($m_1= 0.173$ mag) is low \cite{Rodriguez}. So, this might result in higher luminosity and temperature values, which specify its position on the H-R diagram. Its mass value is also compatible with the theoretical evolution track. Further, since the inclination angle is not known and the rotational velocity is ambiguous, the effects of the rotation on the amplitude spectrum and the evolution status are not clear.

\begin{figure}[!h]
\begin{center}
\includegraphics[width=0.8\textwidth]{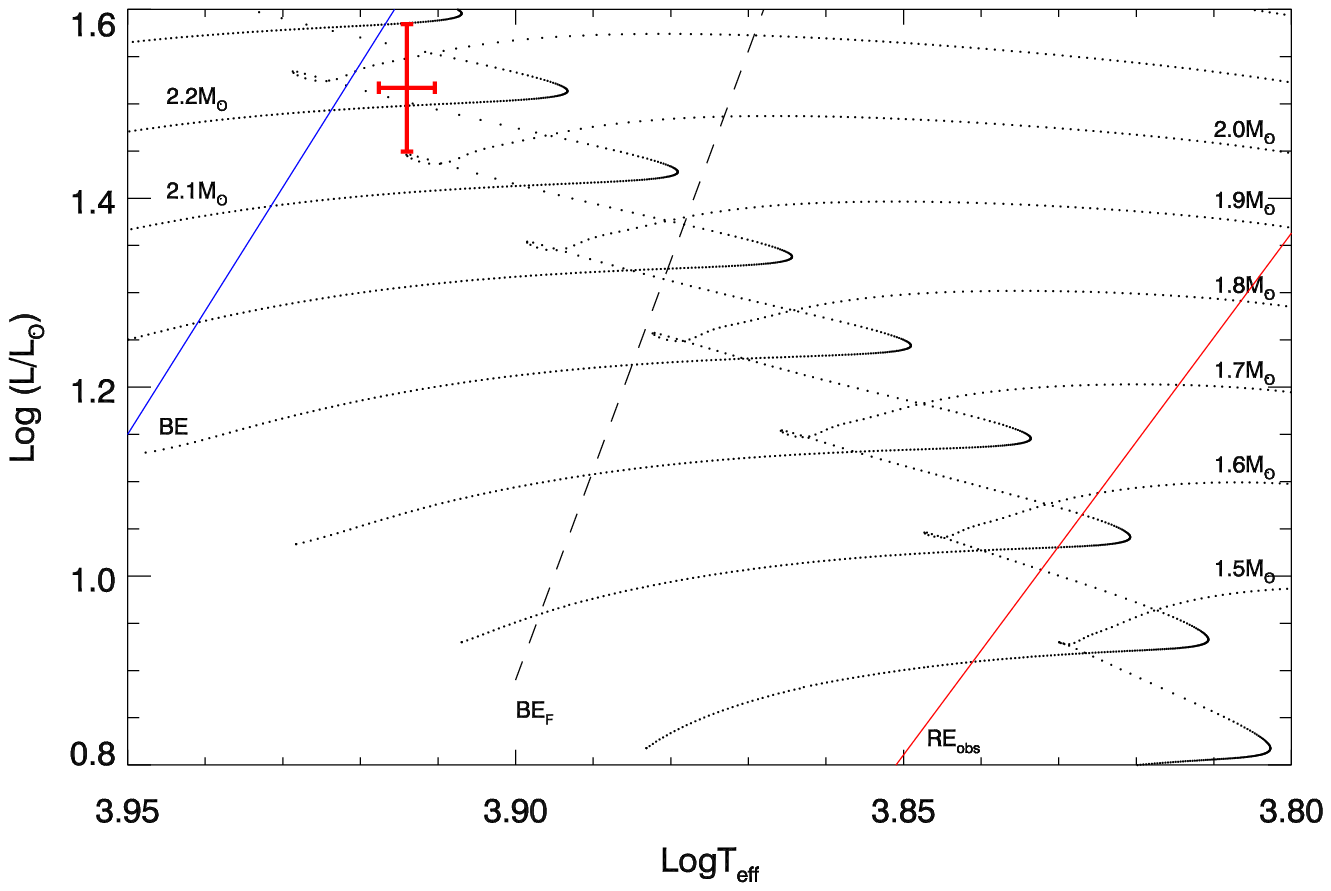}
\caption[HR digram]{The evolutionary stage of HD\,90386. The blue and red lines are the hot and cold edges of the instability strip, respectively. The vertical dashed line is the hot edge for the fundamental mode pulsators (Breger \& Pamyatnykh, 1998b). The theoretical values in the MS and the post-MS phases are taken from Christensen-Dalsgaard \protect\cite{Christensen-Dalsgaard}.}
\end{center}
\label{fig:figure4}
\end{figure}

HD\,90386, a double-mode pulsator, has a simple amplitude spectrum. The period ratio between $f_1$ and $f_3$ is around 0.827, which is quite consistent with the ratio given by Breger \cite{Breger79} for the first and second overtones ($P_2/P_1 = 0.810$). Based on Breger and Bregman \cite{Breger75}, HD\,90386 might be an overtone pulsator since $\delta$ Scuties pulsating with the fundamental mode lie in the cool part of the instability strip whereas the overtone pulsators are found in the hotter part. 

As previously mentioned, the variabilities around 1.00 c~d$^{-1}$ in the amplitude spectrum are consistent with the rotation period. However, there is a possibility that these variations are related to $\gamma$ Dor type high-order g-modes and that HD\,90386 might be a $\delta$ Scuti/$\gamma$ Dor type hybrid star. Although the location of the star on the H-R diagram suggests that it is a $\delta$ Scuti type star, in the studies related {\sl Kepler} A and F type stars Bradley et al. \cite{Bradley} and Uytterhoeven et al. \cite{Uytterhoeven} showed that the locations of $\delta$ Scuti and $\gamma$ Dor classes on the H-R diagram have been extended, and that $\gamma$ Dor type pulsations have been seen in hotter stars than those known so far. They observed a great number of hybrid stars which spread between the blue edge of the $\delta$ Scuti instability strip and the red edge of the $\gamma$ Dor instability strip (two-third of the hybrids have been located in the hot side of the $\delta$ Scuti instability strip \cite{Bradley}. In this context, the frequency accumulation at around 1.00 c~d$^{-1}$ might in fact be related to $\gamma$ Dor type variations and HD\,90386 might be thought as a hybrid star based on the results of these studies.  

Bear in mind that the stellar rotation might also affect the frequency distribution of the star. According to Bouabid et al. \cite{Bouabid}, if the star is a fast rotator, $\gamma$ Dor type pulsations have been shifted to higher frequencies and filled the gap between $\gamma$ Dor type high-order g-modes and $\delta$ Scuti type modes, which is compatible with the frequency range of 5 and 15 c~d$^{-1}$. On this basis, such a situation might explain the low-amplitude variations seen between 5 and 15 c~d$^{-1}$ in the periodogram of the star.

Even though HD\,90386 is classified as a $\delta$ Scuti type pulsator in the literature, it is quite difficult to conclude the actual class of the star by using only the {\sl STEREO} data. Therefore, long-term follow-ups similar to those of {\sl Kepler} satellite, multi-colour photometry, and spectroscopic observations are needed in order to accurately determine physical parameters and the nature of the star.

\end{document}